\documentclass[journal]{IEEEtran}
\pdfoutput=1

\newcommand{\nonubb}  {$0 \nu \beta \beta$}
\newcommand{\twonubb}{$2 \nu \beta \beta$}
\newcommand{\bb}{$\beta\beta$}

\newcommand{\MJ}{{\sc{Majo\-ra\-na}}}
\def\ppc{P-PC}  
\def\nuc#1#2{${}^{#1}$#2}

\def\instUA{$^{1}$}
\def\instBHSU{$^{2}$} 
\def\instUCBNE{$^{3}$}	
\def\instUCB{$^{4}$}	
\def\instUC{$^{5}$}		
\def\instDUKE{$^{6}$}	
\def\instITEP{$^{7}$}
\def\instJINR{$^{8}$}
\def\instLBNLed{$^{9}$}		
\def\instLBNLinpa{$^{10}$}		
\def\instLBNLnsd{$^{11}$}		
\def\instLANL{$^{12}$}			
\def\instUNC{$^{13}$}	
\def\instNCSU{$^{14}$}	
\def\instORNL{$^{15}$}
\def\instOU{$^{16}$}		
\def\instPNNL{$^{17}$}		
\def\instQU{$^{18}$}		
\def\instUSC{$^{19}$}
\def\instUSD{$^{20}$}	
\def\instUT{$^{21}$}		
\def\instTUNL{$^{22}$}	
\def\instUW{$^{23}$}	

\usepackage{graphicx}
\usepackage{epstopdf}

\usepackage{cite}

\begin{document}

\title{The {\sc Majorana} Neutrinoless Double-Beta Decay Experiment}


\author{V.E.~Guiseppe\instLANL,
C.E.~Aalseth\instPNNL,
M.~Akashi-Ronquest\instUNC$^,$\instTUNL,
M.~Amman\instLBNLed,
J.F.~Amsbaugh\instUW,
F.T.~Avignone~III\instUSC$^,$\instORNL,
H.O.~Back\instNCSU$^,$\instTUNL,
A.S.~Barabash\instITEP,
P.~Barbeau\instUC,
J.R.~Beene\instORNL,
M.~Bergevin\instLBNLnsd,
F.E.~Bertrand\instORNL,
M.~Boswell\instUNC$^,$\instTUNL,
V.~Brudanin\instJINR,
W.~Bugg\instUT,
T.H.~Burritt\instUW,
Y-D.~Chan\instLBNLnsd,
T.V.~Cianciolo\instORNL, 
J.~Collar\instUC,
R.~Creswick\instUSC,
M.~Cromaz\instLBNLnsd,
J.A.~Detwiler\instLBNLnsd,
P.J.~Doe\instUW, 
J.A.~Dunmore\instUW,
Yu.~Efremenko\instUT,
V.~Egorov\instJINR,
H.~Ejiri\instOU,
S.R.~Elliott\instLANL, 
J.~Ely\instPNNL,
J.~Esterline\instDUKE$^,$\instTUNL,
H.~Farach\instUSC, 
T.~Farmer\instPNNL,
J.~Fast\instPNNL,
P.~Finnerty\instUNC$^,$\instTUNL,
B.~Fujikawa\instLBNLnsd,
V.M.~Gehman\instLANL,
C.~Greenberg\instUC,
K.~Gusey\instJINR,
A.L.~Hallin\instUA,
R.~Hazama\instOU,
R.~Henning\instUNC$^,$\instTUNL,
A.~Hime\instLANL,
E.~Hoppe\instPNNL,
T.~Hossbach\instUSC$^,$\instPNNL,
M.A.~Howe\instUNC$^,$\instTUNL,
D.~Hurley\instLBNLnsd,
B.~Hyronimus\instPNNL,
R.A.~Johnson\instUW,
K.J.~Keeter\instBHSU,
M.~Keillor\instPNNL,
C.~Keller\instUSD,
J.~Kephart\instNCSU$^,$\instTUNL$^,$\instPNNL,
M.~Kidd\instDUKE$^,$\instTUNL,
O.~Kochetov\instJINR,
S.I.~Konovalov\instITEP,
R.T.~Kouzes\instPNNL,
K.T.~Lesko\instLBNLinpa$^,$\instUCB,
L.~Leviner\instNCSU$^,$\instTUNL,
P.~Luke\instLBNLed,
A.B.~McDonald\instQU,
S.~MacMullin\instUNC$^,$\instTUNL,
M.G.~Marino\instUW,
D.-M.~Mei\instUSD,
H.S.~Miley\instPNNL,
A.W.~Myers\instUW,
M.~Nomachi\instOU,
B.~Odom\instUC,
J.~Orrell\instPNNL,
A.W.P.~Poon\instLBNLnsd,
G.~Prior\instLBNLnsd,
 D.C.~Radford\instORNL,
J.H.~Reeves\instPNNL,
K.~Rielage\instLANL,
N.~Riley\instUC,
R.G.H.~Robertson\instUW, 
L.~Rodriguez\instLANL,
K.P.~Rykaczewski\instORNL,
A.G.~Schubert\instUW,
T.~Shima\instOU,
M.~Shirchenko\instJINR,
J. Strain\instUNC$^,$\instTUNL,
R.~Thompson\instPNNL,
V.~Timkin\instJINR,
W.~Tornow\instDUKE$^,$\instTUNL,
C.~Tull\instLBNLnsd,
T. D.~Van Wechel\instUW,
I.~Vanyushin\instITEP,
R.L.~Varner\instORNL,
K.~Vetter\instLBNLnsd$^,$\instUCBNE,
R.~Warner\instPNNL,
J.F.~Wilkerson\instUNC$^,$\instTUNL,
J.M.~Wouters\instLANL,
E.~Yakushev\instJINR,
A.R.~Young\instNCSU$^,$\instTUNL,
C.-H.~Yu\instORNL,
V.~Yumatov\instITEP,
C.~Zhang\instUSD\\
(\MJ\ Collaboration)


%
\thanks{\instUA\ Centre for Particle Phys., Univ. of Alberta, Edmonton, Alberta, Canada}%
\thanks{\instBHSU\ Department of Physics, Black Hills State University, Spearhead, SD, USA}
\thanks{\instUCBNE\ Department of Nuclear Engineering, UC Berkeley, Berkeley, CA, USA}
\thanks{\instUCB\ Department of Physics, UC Berkeley, Berkeley, CA, USA}
\thanks{\instUC\ University of Chicago, Chicago, IL, USA}
\thanks{\instDUKE\ Department of Physics, Duke University, Durham, NC, USA}
\thanks{\instITEP\ Institute for Theoretical and Experimental Physics, Moscow, Russia}
\thanks{\instJINR\ Joint Institute for Nuclear Research, Dubna, Russia}
\thanks{\instLBNLed\ Engineering Div., Lawrence Berkeley Natl. Lab., Berkeley, CA, USA}
\thanks{\instLBNLinpa\ Inst. for Nucl. \& Part. Astro., Lawrence Berkeley Natl. Lab., Berkeley, CA, USA}
\thanks{\instLBNLnsd\ Nuclear Science Div., Lawrence Berkeley Natl. Lab., Berkeley, CA, USA}
\thanks{ \instLANL\ Physics Div., Los Alamos National Laboratory, Los Alamos, NM, USA}
\thanks{\instUNC\  Dept. of Phys. and Astron., Univ. of North Carolina, Chapel Hill, NC, USA}
\thanks{\instNCSU\ Department of Physics, North Carolina State Univ., Raleigh, NC, USA}
\thanks{\instORNL\ Oak Ridge National Laboratory, Oak Ridge, TN, USA}
\thanks{\instOU\ Res. Center for Nucl. Phys. \& Dept. of Phys., Osaka Univ., Ibaraki, Osaka, Japan}\thanks{\instPNNL\ Pacific Northwest National Laboratory, Richland, WA, USA}
\thanks{\instQU\ Dept. of Physics, Queen's University, Kingston, Ontario, Canada}
\thanks{\instUSC\ Dept. of Phys. and Astron., Univ. of South Carolina, Columbia, SC, USA}
\thanks{\instUSD\ Dept. of Earth Science and Phys., Univ. of South Dakota, Vermillion, SD, USA}
\thanks{\instUT\ Dept. of Physics and Astron., Univ. of Tennessee, Knoxville, TN, USA}
\thanks{\instTUNL\ Triangle Universities Nuclear Laboratory, Durham, NC, USA}
\thanks{\instUW\ Center for Nucl. Phys. and Astrophys., Univ. of Washington, Seattle, WA, USA}
\thanks{V.E. Guiseppe is with Los Alamos National Laboratory, Los Alamos, NM 87545 USA (e-mail: guiseppe@lanl.gov)}

}


\maketitle
\thispagestyle{empty}

\begin{abstract}
Neutrinoless double-beta decay searches play a major role in
determining the nature of neutrinos, the existence of a lepton violating process, and  the effective Majorana neutrino mass. The M{\sc
ajorana} Collaboration proposes to assemble an array of HPGe
detectors to search for neutrinoless double-beta decay in
$^{76}$Ge. 
Our proposed method uses the well-established technique of searching for neutrinoless double-beta decay in high purity
Ge-diode radiation detectors that play both roles of source and detector. The technique
is augmented with recent improvements in signal processing and detector design, and advances in
controlling intrinsic and external backgrounds.
Initially, M{\sc ajorana} aims to construct a
prototype module containing 60 kg of Ge detectors to demonstrate
the potential of a future 1-tonne experiment. The design and
potential reach of this prototype Demonstrator module will be presented. This
paper will also discuss detector optimization and low-background requirements, such as material purity, background rejection, and identification of rare backgrounds required to reach the sensitivity goals of the M{\sc ajorana}
experiment. 

\end{abstract}


\section{Introduction}
\IEEEPARstart{W}{ith} the realization that neutrinos are not massless particles, but have mass and mix, there is increased
interest in investigating their intrinsic properties.  Understanding
the neutrino mass generation mechanism, the absolute neutrino mass
scale and the neutrino mass spectrum are some of the main focuses
of future neutrino experiments.
The discovery of Majorana neutrinos would have profound
theoretical implications in the formation of a new Standard Model
while yielding insights into the origin of mass itself.
As of yet, there is no firm experimental
evidence to confirm or refute this theoretical prejudice.
Experimental evidence of neutrinoless double-beta (\nonubb) decay
would definitively establish the Majorana nature of neutrinos.
Not only is
\nonubb\  decay the only practical method to uncover
the Majorana nature of neutrinos, it has the potential to reach an absolute mass scale sensitivity of $<$50~meV.
Observation of a sharp peak at the \ensuremath{\beta}\ensuremath{\beta} endpoint
would quantify the \nonubb-decay rate,
demonstrate that neutrinos are Majorana particles, indicate that
lepton number is not conserved, and provide a measure of the
effective Majorana mass of the electron neutrino \cite{ell02, ell04, bar04, avi05, eji05, avi08}. 

Ordinary beta decay of many heavy even-even nuclei is energetically
forbidden or strongly spin inhibited. However, a process in which a nucleus changes its atomic
number by two while simultaneously emitting two beta particles
is energetically possible for some even-even nuclei. In two-neutrino double-beta decay
(\twonubb), the nucleus emits 2 $\beta$ particles and 2 $\overline{\nu } _{e}$ conserving lepton number.
It is an allowed second-order weak process that occurs in
nature, although its rate is extremely low. Half-lives for this
decay mode have been measured at $\sim$10$^{19}$ years
or longer in several nuclei~\cite{bar06}.

The more interesting process is zero-neutrino double-beta decay
(\nonubb) where only the 2 $\beta$ particles are emitted and no neutrinos.
Unlike \twonubb,
\nonubb\ violates
lepton number conservation and hence requires physics beyond the
Standard Model. One can visualize
\nonubb\ as an exchange
of a virtual neutrino between two neutrons within the nucleus. In
the framework of the 
SU$_{L}(2) \times$ U$(1)$ Standard Model of weak interactions, the first
neutron emits a right-handed anti-neutrino. However, the second
neutron requires the absorption of a left-handed neutrino. In order
for this to happen, the neutrino must have mass so that it is not in
a pure helicity state, and the neutrino and anti-neutrino have
to be indistinguishable. That is, the neutrino would have to be a
massive Majorana particle.

One isotope known to undergo \bb-decay is $^{76}$Ge.
The \MJ\ Collaboration has proposed an experiment \cite{ell08} using the well-established technique
of searching for \nonubb\ decay in high-purity
Ge-diode radiation detectors that play both roles of source and
detector. The technique maximizes the source to total mass ratio and benefits from excellent energy resolution (0.16\% at 2.039 MeV). Ge detectors are able to be enriched in the \bb-decay isotope $^{76}$Ge from 7.44\% to 86\%.
Ge-based \nonubb\ experiments have established the best half-life limits and the most restrictive constraints on the effective Majorana mass for the neutrino \cite{aal02a,bau99}. One analysis of the data in Ref. \cite{bau99} claims evidence for \nonubb\ with a half-life of $2.23^{+0.44}_{-0.31} \times 10^{25}$ y \cite{kla06}.

The future use of Ge detectors can be augmented with recent improvements in
signal processing and detector design, and advances in controlling
intrinsic and external backgrounds.  Progress
in signal processing from segmented Ge-diode detectors 
potentially offers
significant benefits in rejecting backgrounds, reducing sensitivity
of the experiment to backgrounds, and providing additional handles
on both signals and backgrounds through multi-dimensional event
reconstruction.  Development of sophisticated Cu-electroforming
methods allow the fabrication of ultra-low-background materials
required for the construction of next-generation experiments.

\section{The \MJ\ Approach}
The \MJ\ collaboration is currently pursuing R\&D aimed at a 1-tonne scale,
$^{76}$Ge \nonubb-decay experiment that potentially would be one of the initial suite of
experiments to be sited at the U.S. Deep Underground Science and Engineering Laboratory (DUSEL) in the former Homestake Mine in South Dakota \cite{dusel}. The goals of a 1-tonne experiment would be to determine the Majorana or Dirac nature of neutrinos, test lepton number conservation, and probe the absolute neutrino mass scale at the atmospheric mass scale of 20-40 meV. We are currently cooperating with the European GERDA  Collaboration \cite{sch05} with the aim to jointly prepare for a single international tonne-scale Ge-based experiment utilizing the best technologies of the two collaborations.

\begin{figure}
\centering
\includegraphics[width=3.5in]{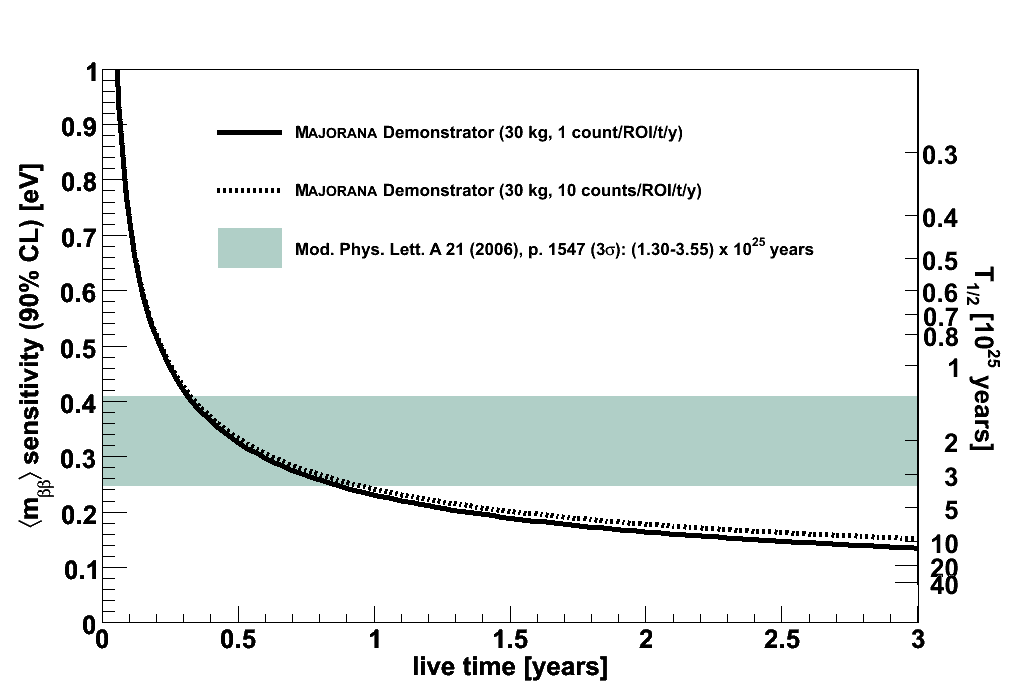}
\caption{The \nonubb\ half-life and effective Majorana mass sensitivity of the \MJ\ Demonstrator module.}
\label{fig:dem-sens}
\end{figure}

For the R\&D phase, the \MJ\ collaboration intends to construct
a Demonstrator module of $^{76}$Ge crystals contained
in an ultra-low-background structure.  The \MJ\ R\&D Demonstrator module should 
significantly improve the lower limits on the decay lifetime from the current
level of about $2 \times 10^{25}$ years to about $7 \times 10^{26}$
years, corresponding to an upper limit of 90~meV on the effective Majorana electron-neutrino mass (Fig. \ref{fig:dem-sens}). The Demonstrator module will be located at the 4850-ft level (4200 m.w.e) at the Sanford Laboratory, the future home of DUSEL.

\begin{figure}
\centering
\includegraphics[width=2.5in]{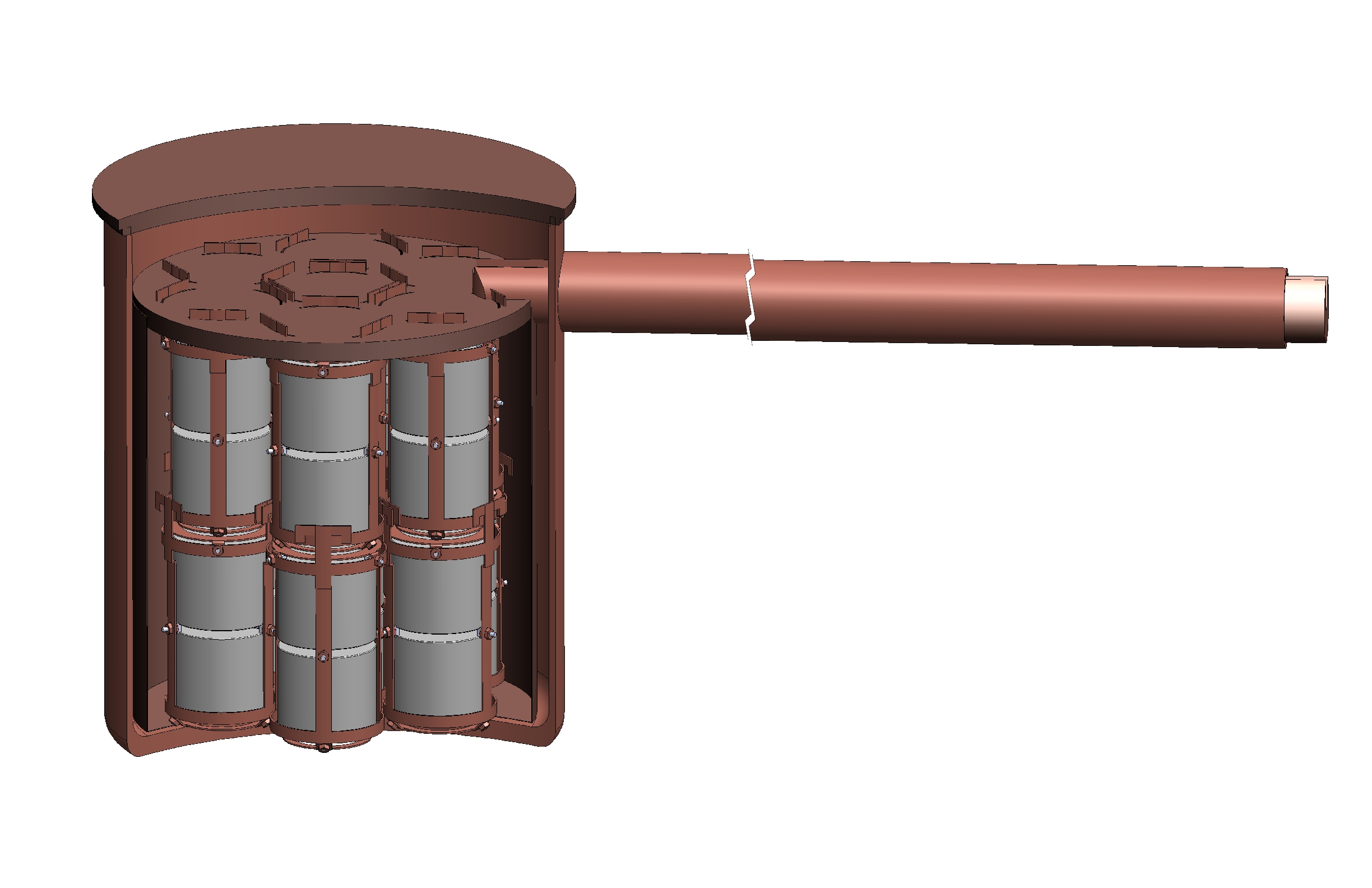}
\caption{A representation of the one of three low-mass, low-background cryostat design of the \MJ\ Demonstrator.}
\label{fig:cry}
\end{figure}

The  Demonstrator reference design is a modular concept with three individual 
cryostats (Fig. \ref{fig:cry}) containing a combination of enriched (86\%-enrichment) and unenriched
(natural) Ge crystals. This R\&D system allows exploration of segmentation
schemes and detector types. Our initial emphasis will be on a first cryostat assembled with natural-Ge \ppc\ detectors. The modular concept provides timely
step-wise installation, access, and future up-scaling.  By using 30 kg of 86\% enriched $^{76}$Ge crystals, the \nonubb\ claim \cite{kla06} will be tested after 2 y of running. The additional 30 kg of natural Ge allows enough total sensitivity to understand backgrounds. The background goal in the \nonubb\ peak 4-keV region of interest centered at 2039 keV is 1 count/tonne/year after analysis cuts.

The detectors will
be mounted in a string-like arrangement as shown in Figure~\ref{fig:cry}.
The reference design of the string, which continues to be refined, currently consists of a
thick copper ``lid'', a copper support frame, and low-mass plastic
support trays or stand-offs.  Each individual detector is mounted
in a separate frame built of electro-formed Cu that eases handling
of individual detectors during shipping, acceptance testing and
string assembly.  The assembled detector string is lowered
through a hole in the cryostat cold plate until the string lid sits
on the cold plate.  This allows us to mount and dismount
individual detector strings from the top. 
Cables are run vertically
from the detector contacts through a slit in the string lid, above
which low-background front-end electronics packages are mounted for
reading out the central contact. 
The string lid
provides some shielding between these components and the detectors,
and allows the bandwidth to be maximized by placing the central
contact FETs near the Ge detectors.  

All crystals are
enclosed in ultra-pure, electro-formed copper cryostats.  A thermal shroud mounted to the
cold plate provides radiative cooling for the detector array.
The cryostat, cold plate and thermal shroud will be fabricated using
ultra-low-back\-ground electro-formed copper.  This copper is the lowest 
radioactivity
material we will have available and a relatively thick cryostat
wall can be used to make a strong
vacuum vessel.  See Section \ref{sect:bg} for more details on the electro-forming process.

\begin{figure}
\centering
\includegraphics[width=2.5in]{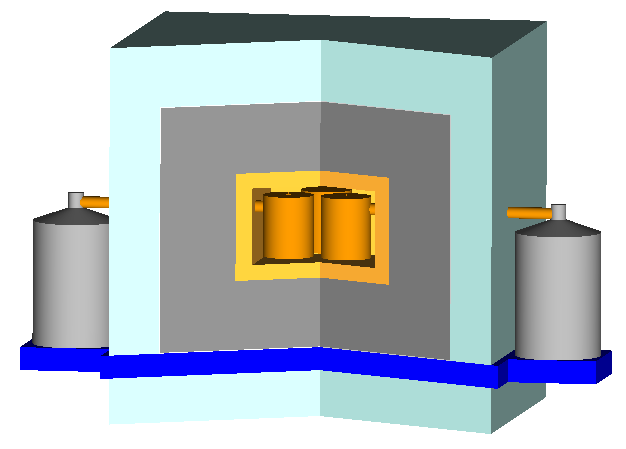}
\caption{A representation of the passive shield for the \MJ\ Demonstrator. Shown are three cryostats surrounding by Cu, Pb, and neutron moderating passive shielding layers.}
\label{fig:shield}
\end{figure}

The cryostats will be enclosed in a graded passive shield (Fig. \ref{fig:shield}) and
an active muon veto to eliminate external backgrounds.
Shielding reduces signals from
$\gamma$ rays originating in the experiment hall (from rock,
construction materials, and from the shielding materials
themselves), cosmic-ray $\mu$'s penetrating the shielding, and
cosmic-ray $\mu$-induced neutrons. The strategy is to provide
extremely low-activity material for the {\em inner shield}. Surrounding
this will be an {\em outer shield} of bulk $\gamma$-ray shielding material 
with lower radiopurity.  This high-$Z$ shielding
enclosure will be contained inside a gas-tight Rn exclusion box
made of stainless-steel sheet. Outside this bulk high-$Z$ shielding
will be a layer of hydrogenous material, some of which will be doped
with a neutron absorber such as boron, intended to reduce the neutron
flux.  Finally, active cosmic-ray anti-coincidence detectors
will enclose the entire shield. 

Monte Carlo (MC) radiation transport simulation models have been
developed for \MJ\ using MaGe \cite{bau06}, an object-oriented MC simulation
package based on ROOT~\cite{bru97} 
and the Geant4~\cite{ago03, all06} toolkit. MaGe is 
optimized for simulations
of low-background Ge detector arrays and is being jointly
developed by the \MJ\ and GERDA~\cite{sch05} collaborations.
MaGe defines a set of physics processes, materials, constants,
event generators, etc.~that are common to these experiments, and
provides a unified framework for geometrical definitions, database
access, user interfaces, and simulation output schemes in an effort
to reduce repetition of labor and increase code scrutiny.

Typical HPGe digitization electronics entail a digitizing ADC read
out by an on-board FPGA. Such a card has been custom-designed to
readout the GRETINA detectors \cite{des05}. Such a board is well-matched to
the digital signal processing needs of \MJ. The DAQ software system will be constructed using the Object-oriented
Real-time Control and Acquisition (ORCA)~\cite{how04} application which provides a general purpose, highly modular, object-oriented,
acquisition and control system that is easy to develop, use, and
maintain. With the ability to control embedded single board computers, ORCA extends its ease of use to a wide range of hardware options \cite{how08}.

\section{Backgrounds}
\label{sect:bg}
Mitigation of backgrounds is crucial to the success of any rare
decay search. For the case of Ge solid-state
detectors, decades of research have yielded a host of techniques to
reduce backgrounds. These techniques include the use of ultra-pure
materials for the construction of detector components in the
proximity of the crystals, shielding the detectors from external
natural and cosmogenic sources, and optimizing detector energy
resolution to enhance the spectral information available. The key to the \MJ\ design is the ability to reduce backgrounds to unprecedented levels. 

Typical backgrounds that require mitigation include the natural radioactive chains, anthropogenic isotopes, and cosmogenics. The isotopes produced by cosmic activation are particularly problematic. At the Earth's surface, cosmic rays can activate isotopes in detector materials. These backgrounds will be reduced by carefully limiting exposure of detector and susceptible shielding materials above ground. As much production as possible will take place underground. Once underground, backgrounds can be produced from the remaining muon-induced hard, secondary neutrons  interacting in the detector and shielding materials. The detector shielding will utilize neutron moderating material and an active muon veto to limit this class of background. 

Copper is one of the very few elements having no relatively
long-lived radioisotopes. It also has excellent physical, chemical, and
electronic properties that make it particularly useful in the
fabrication of low-background radiation detectors. Nevertheless, one
must take care to ensure that it is not contaminated with
radioactive impurities, and that it does not have significant
quantities of cosmic ray generated radioisotopes. The most serious
of these latter impurities is $^{60}$Co, which is generated by
(n,$\alpha$) reactions on $^{63}$Cu. Ultra-pure copper
for use in \MJ\ will be produced by an electro-forming process underground. 

The electro-forming process \cite{hop07} uses an acidic CuSO$_{4}$ solution through
which positive copper ions are induced from a copper anode to a
stainless steel mandrel which initially serves as the cathode. The
sacrificial copper anode material is consumed and replaced as
necessary as the electro-formed part is formed at the cathode.
Highest purity copper is obtained by carefully limiting the
electro-forming potential to selectively eliminate a wide variety
of contaminants that reside in the bath as the less pure anode
copper source material is consumed.  Optimal material is achieved by controlling
the rate of plating (i.e.~current density at the Cu surface),
manipulation of electrode surface boundary layers (via agitation,
reverse pulse plating, etc.), and the use of crystal growth inhibiting
chemicals.  The optimization of electro-forming copper with adequate
structural integrity and high purity is underway.

Copper cleanliness studies have shown that the CuSO$_{4}$ in the bath is the source of Th in the produced copper parts.  Producing CuSO$_{4}$ from pure starting materials has been more successful in producing clean copper parts than re-cystalizing the CuSO$_{4}$.

Recent
advances have provided new background-rejection techniques in the form of pulse-shape
analysis, detector segmentation, and granularity
(inter-detector coincidences). These techniques rely on the different
spatial distributions of energy deposition between
double-beta decay events and most background signals.
Background signals arising from radioactive decay often include
a beta and one or more $\gamma$ rays. In addition, a 2-MeV 
$\gamma$ ray frequently undergoes multiple scatters on the centimeter
scale.  Double-beta decay energy deposition occurs within a
small volume ($<$1~mm$^{3}$) and  hence is single site.

The pulse-shape analysis makes use of a digitized preamplifier pulse shape 
to extract key pulse parameters, including the width, asymmetry, kurtosis, and
higher moments. Analysis of these parameters indicate whether an event is
single- or multi-site. Electrically subdividing the detector into smaller elements gives
additional segmentation discrimination between single site events like \nonubb\
decay and internal or external $\gamma$-ray backgrounds.
In a closely packed array of detectors, there is a large probability
for hits in multiple detectors from internally generated
$\gamma$ rays that escape a crystal, Compton scatter
$\gamma$ rays from external sources, and traversing
muons. The appearance of hits in several nearby detectors within a
short (few microsecond) coincidence window provides a
large-scale multiplicity or granularity cut. Single-site, time-correlated (SSTC) cuts 
are decay-chain-specific. This method looks forward or
backwards in time from an event in the ROI to find signatures of
parent or daughter isotopes~\cite{eji05}. Given a raw event
rate of roughly 1~event per crystal per day, this method will work
exceptionally well for internal, short-lived parent-daughter pairs like
$^{68}$Ge-$^{68}$Ga and for decays of internal contaminants, with
somewhat lower rejection efficiency. It will also play an important
role in diagnosing and eliminating surface contamination of U and
Th chain isotopes.

 \begin{figure}
 \centering
\includegraphics[width=3in]{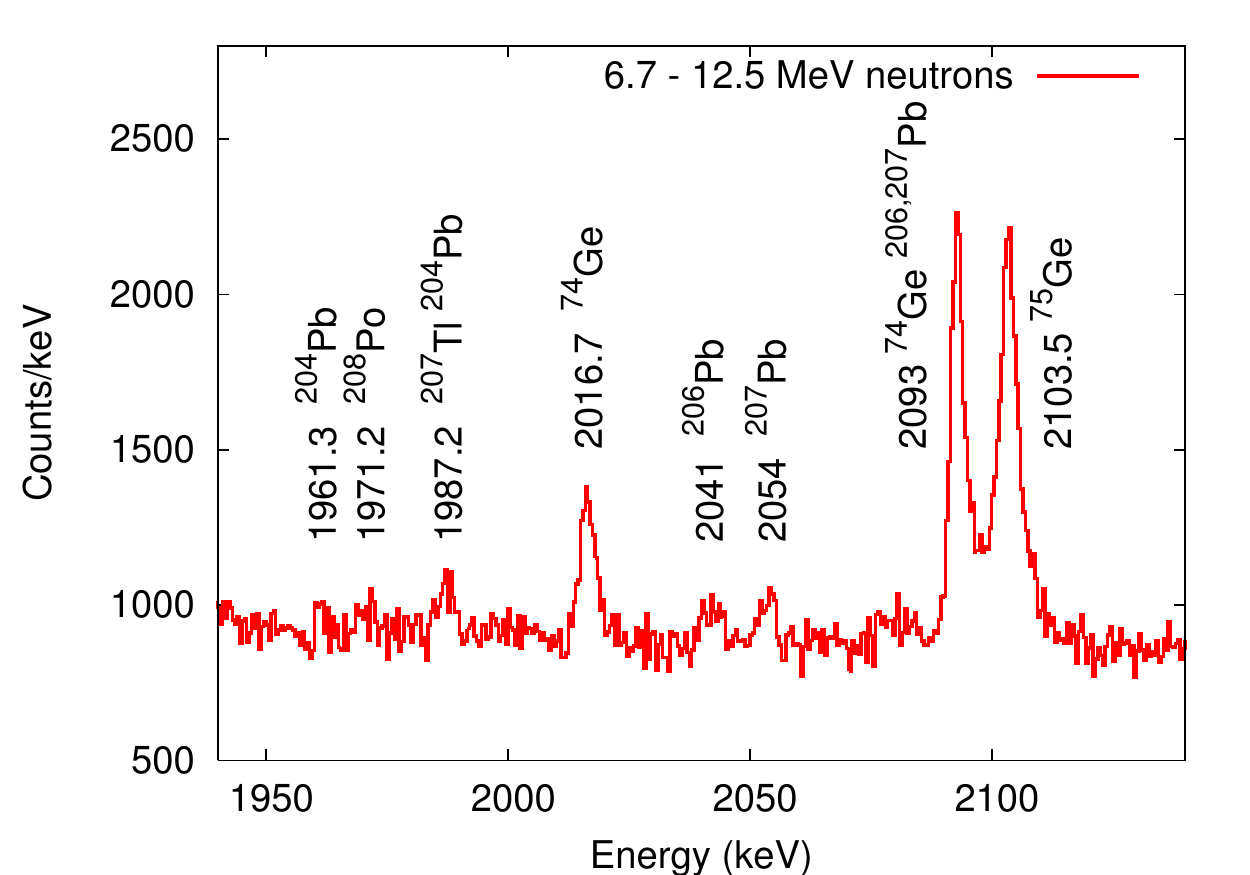}\\
\includegraphics[width=3in]{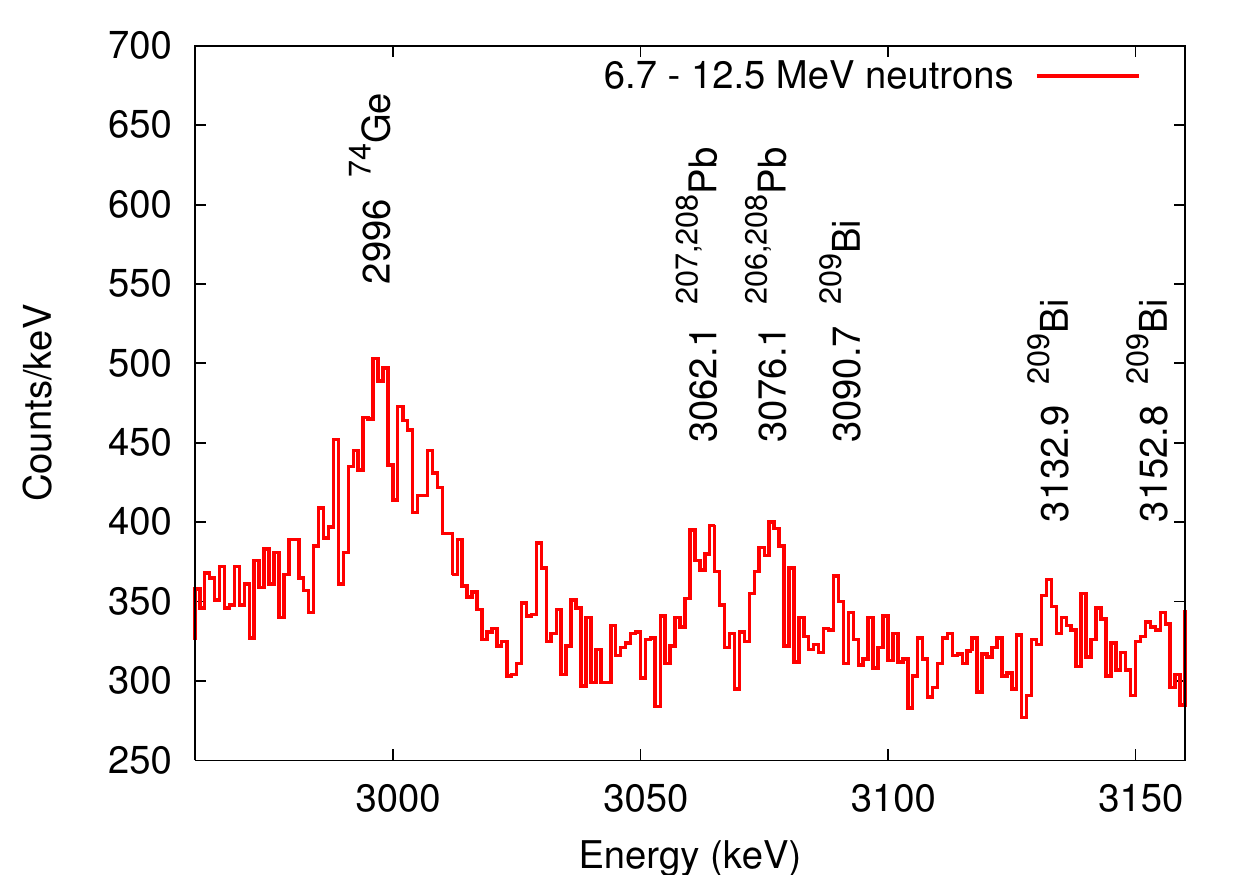}
\caption{The $\gamma$-ray spectra showing the 2041-keV $\gamma$-ray line and the 3062-keV $\gamma$-ray line from 6.7-12.5 MeV neutrons \cite{gui08}.}
 \label{fig:pb-lines}
\end{figure}

If the efforts to reduce the natural decay chain isotopes are successful, previously unimportant components of the background must be understood and eliminated. For example, $(n,n'\gamma)$ reactions will become important for tonne-scale double-beta decay experiments \cite{mei06}. Specific Pb $\gamma$-rays are problematic backgrounds for Ge-based  experiments using large quantities of Pb shielding: \nuc{206}{Pb} has a 2040-keV $\gamma$ ray, \nuc{207}{Pb} has a 3062-keV $\gamma$ ray, and \nuc{208}{Pb} has a 3060-keV $\gamma$ ray. The first $\gamma$ ray is near the \nuc{76}{Ge} \bb\ Q-value of 2039 keV. The double escape peak of the latter two is a single-site energy deposit near the \bb\ Q-value. The levels that produce these $\gamma$ rays can be excited by (n,n$^\prime \gamma$) or (n,xn$\gamma$) reactions, but the cross sections are small and previously unmeasured. Excitation of these levels (Fig. \ref{fig:pb-lines}) were studied with a broad-energy neutron beam to measure reaction cross sections \cite{gui08}. The cross sections can be folded with the underground neutron flux to estimate background rates.

\section{Detectors}
\begin{figure}
\centering
\includegraphics[width=3.5in]{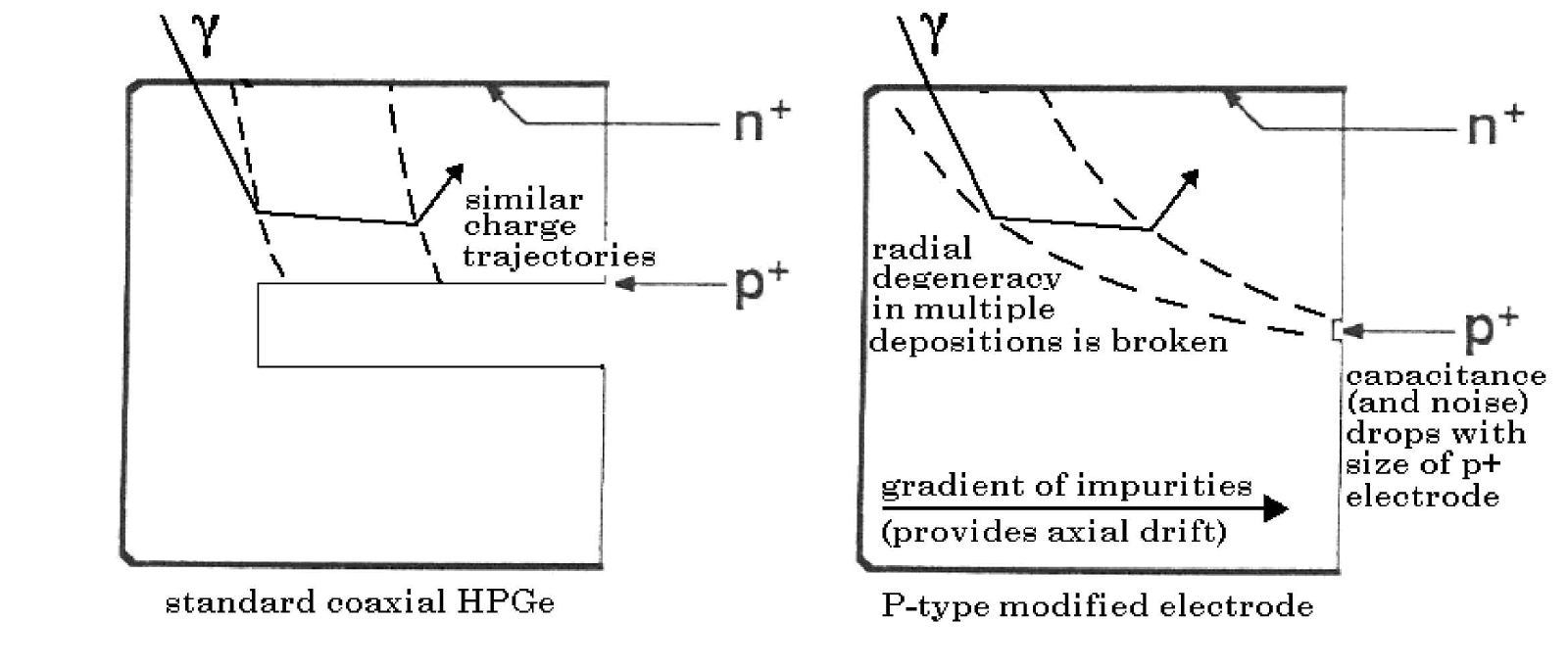}
\caption{Closed-end coaxial p-type Ge detector (left) versus 
the \ppc\ Ge detector implementation (right) \cite{bar07}.}
\label{fig:pmeopt}
\end{figure}

We are  exploring two Ge detector technologies. The first are novel point-contact, p-type (\ppc) detectors (Fig. \ref{fig:pmeopt}),
which  have an excellent interaction separation capability, superior energy
resolution and low energy threshold. This is due to the point
contact's small capacitance and the rapidly changing electrical
weighting field in the vicinity of the readout contact.
This approach provides a very effective active
background separation by pulse-shape analysis with only one readout
channel and therefore minimum amount of potentially radioactive
materials, particularly close to the detector.

\begin{figure}
\centering
\includegraphics[width=3in]{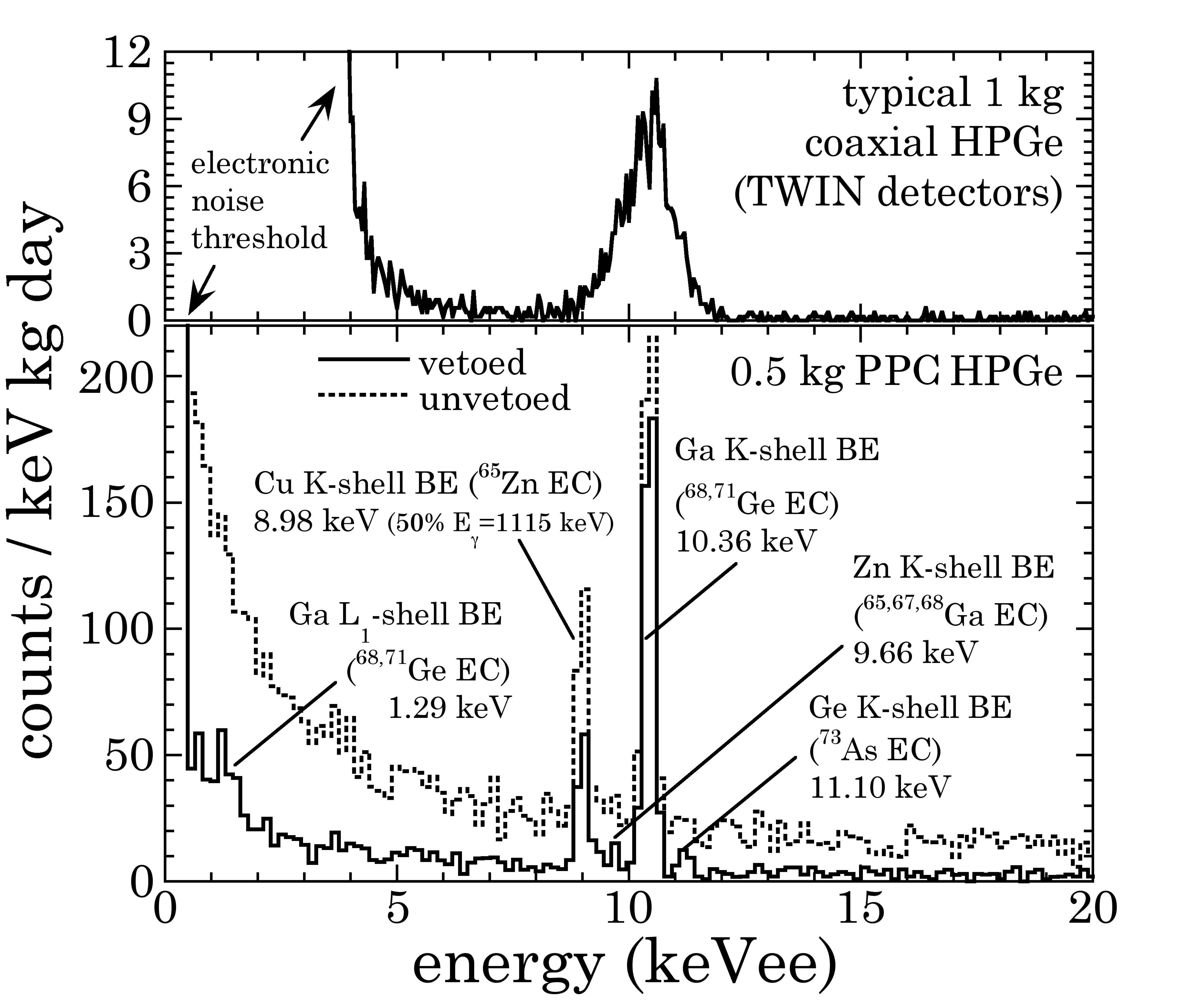}
\caption{A demonstration of the improved energy resolution and threshold for \ppc\ detectors versus a typical coaxial HPGe detector \cite{aal08}.}
\label{fig:ppcthreshold}
\end{figure}

\begin{figure}
\centering
\includegraphics[width=2.5in]{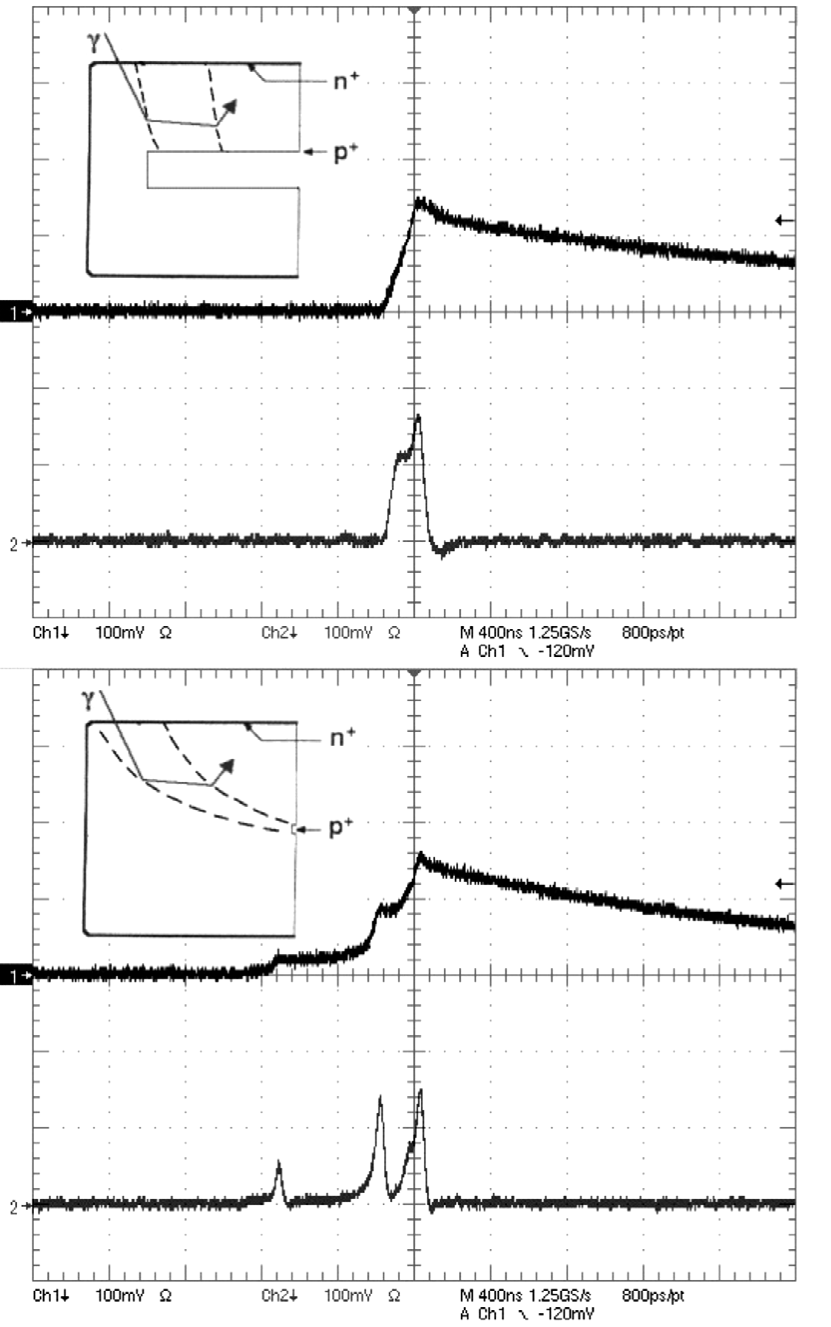}
\caption{Pulse shape from closed-end coaxial p-type Ge detector (top) versus 
the \ppc\ Ge detector implementation (bottom) \cite{bar07}.}
\label{fig:ppcdrift}
\end{figure}

\begin{figure}
\centering
\includegraphics[width=2.5in]{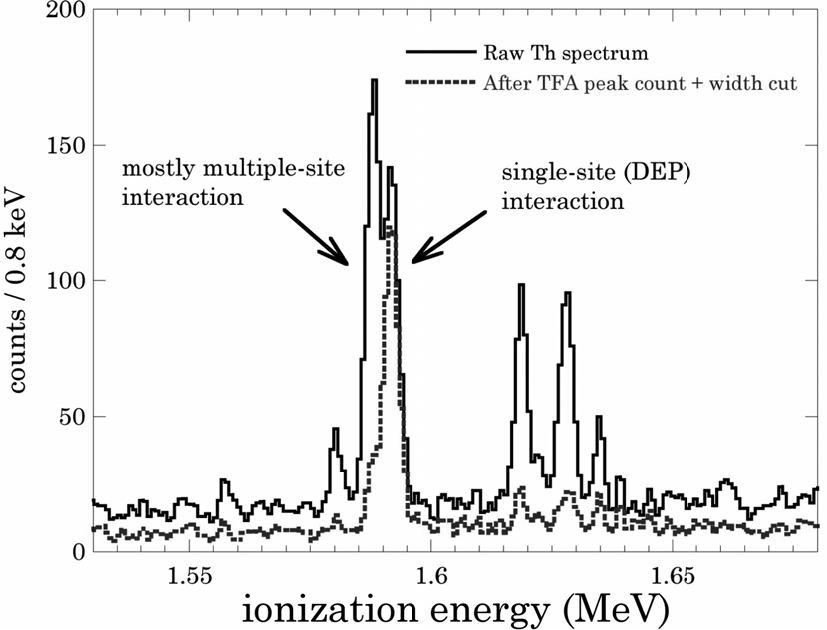}
\caption{Pulse shape analysis  of a \ppc\ Ge detector showing the efficiency of rejecting multi-site events (high-energy $\gamma$ rays) and accepting single-site events (a double escape peak) \cite{bar07}.}
\label{fig:psa}
\end{figure}

The \ppc\ is an alternative right circular cylindrical detector design 
developed in which the bore hole is removed and replaced with a
point-contact, either implanted B or drifted Li, in
the center of the passivated detector face~\cite{luk86}.  The
changes in the electrode structure result in a drop in capacitance
to $\sim$1~pF, reducing the electronics noise component and enabling
sub-keV energy thresholds.  This \ppc\
configuration also has lower electric fields throughout the bulk
of the crystal and a weighting potential that is sharply peaked
near the point contact. This in turn results in an increased range
of drift times and a distinct electric signal marking the arrival
of the charge cloud at the central electrode.  Figure~\ref{fig:pmeopt}
illustrates the point-contact implementation in contrast to the
conventional, closed-end coaxial detector approach.  Instrumented
with a modern FETs, \ppc\ detectors have recently been demonstrated by
\MJ\ collaborators to provide low noise, resulting in a low energy
trigger threshold and excellent energy resolution (Fig. \ref{fig:ppcthreshold}), as well as
excellent pulse-shape capabilities (Figs. \ref{fig:ppcdrift} and \ref{fig:psa}) to distinguish multiple
interactions~\cite{bar07}.

Beyond background rejection capabilities, \ppc\ detectors also have
potentially increased manufacturing speed, lower cost, simpler
construction and data analysis, and decreased thermal load and
photon path.  It is anticipated that several crystal growers and vendors will
 potentially be able to provide this type of detector.  Handling
and characterization of these detectors is straight forward.
Several
measures are under study that would reduce surface contamination
backgrounds to negligible levels for this type of device: pairing
detectors with their passivated ends face-to-face (the only surfaces
sensitive to alphas in these devices), and (separately) possibly
detecting particle induced x-ray emission from alphas entering
one of them.

Until recently, only one operational \ppc\ detector had been produced,
and the chief uncertainty with the \ppc\ approach was the requirement
for the impurity concentration profile in the crystal. During the last
year, however, five additional detectors have been produced by three
different vendors or institutions (Canberra, Lawrence Berkeley National Laboratory, and PHD's Inc.).
In
addition to demonstrating that \ppc\ detectors are not difficult to
produce, the new detectors have validated field calculation codes and
simulations that allow us to determine the impurity gradient
requirements and the dimensions to which such detectors can reliably
be manufactured.  This information also enables simulations to design optimal detectors for \MJ.

 The second detector technology is a segmented detector approach based on large-volume
n-type detectors. This approach not
only provides the discrimination of single and multiple interactions
in three dimensions, it also provides the ability to fully reconstruct
individual $\gamma$ rays by measuring the three-dimensional position
(to 2~mm)
and energies of each interactions point.  Such a capability
provides additional information to distinguish the signal from potential
backgrounds. 

Detectors with a moderate degree of segmentation
offer some of the event characterization capability of highly
segmented detectors, but with lower complexity, risk, and cost.
The segment multiplicity provides an additional handle to separate
signal from background over the radial pulse shape analysis and
detector granularity techniques of unsegmented p-type detector arrays.
In addition, by employing pulse-shape analysis of the segment signals,
an improved position sensitivity can be obtained, depending on the
number and orientation of the segmentation.  Simulations and recent
experiments have shown that already with four segments a $\gamma$ ray
of about 2~MeV that leaves its full energy in a detector can be
suppressed by a factor of $\sim$3 just by employing the segment
multiplicity.  Employing 40-fold segment multiplicity,
the suppression factor only increases to about 5.
Compared to non-segmented designs, additional
contacts and cables will be required for the segment signals,
increasing the risk of extra background sources close to the
detectors. 

The Segmented Enriched-Germanium Assembly (SEGA) prototype detector is a 12-segment n-type Ge detector enriched to 86\% in $^{76}$Ge. It has been used for segmentations studies and pulse-shape analysis techniques. Work is underway to re-mount the crystal in a low background  cryostat to study backgrounds. The detector will be operated underground.

\section{Conclusion}
\MJ\ is conducting R\&D for a 1-tonne \nonubb\ decay experiment by 
constructing a Demonstrator module of $^{76}$Ge crystals contained
in an ultra-low-background structure.  Our current Demonstrator reference design is a modular concept with three individual 
cryostats containing a combination of 30 kg of enriched (86\%-enrichment) and 30 kg of unenriched
(natural) Ge crystals.  This amount of material should demonstrate that the backgrounds are low enough to justify scaling to 1-tonne, as well as probing the neutrino effective mass region above 100 meV.

\section*{Acknowledgment}

Preparation of this paper was supported by the U.S. DOE Office of Nuclear Science, the Los Alamos Laboratory Directed Research and Development, and the U.S. NSF Particle and Nuclear Astrophysics.




\bibliographystyle{ieeetr}
\bibliography{/Users/vguiseppe/work/latex/bib/mymj_ieee}
\end{document}